\documentclass{article}
\usepackage{amssymb}

\usepackage{graphicx}
\usepackage{amsmath}

\input{tcilatex}

\begin{document}

\title{Construction of bilinear control Hamiltonians using the series product and
quantum feedback }
\author{J. Gough\thanks{\textit{email:} jug@aber.ac.uk} \\
Institute for Mathematical and Physical Sciences, \\
Aberystwyth University, Ceredigion, SY23 3BZ, Wales}
\date{}
\maketitle

\begin{abstract}
We show that it is possible to construct closed quantum systems governed by
a bilinear Hamiltonian depending on an arbitrary input signal. This is
achieved by coupling the system to a quantum input field and performing a
feedback of the output field back into the system to cancel out the
stochastic effects, with the signal being added to the field between these
events and later subtracted. Here we assume the zero time delay limit
between the various connections and operations.
\end{abstract}

\section{Introduction}

The theory of quantum control can be divided into open loop and closed loop
problems. In open loop problems a common situation is to investigate various
notions of controllability for closed systems governed by a bilinear
Hamiltonian, that is, taking the form 
\begin{equation}
H\left( \mathbf{u}\right) =H_{0}+\sum_{k}H_{k}u_{k}
\end{equation}
where the $H_{0},H_{k}$ are self-adjoint and the $u_{k}$ are real-valued
functions of time called the controls \cite{D'Alessandro},\cite{Sch01}.
Closed loop problems can be divided into two classes: \textit{%
measurement-based control} which requires feedback from a controller based
on quantum state filtering using the results of a measurement of the output 
\cite{Belavkin83}-\cite{BvH}; \textit{coherent control} where the feedback
is fully quantum and no measurement is performed \cite{Lloyd}-\cite{Mabuchi}%
. Typically in closed loop problems, the information is carried from system
to apparatus/controller by quantum input processes \cite{Gardiner Collett}
and so the system undergoes a stochastic open dynamical evolution. A theory
for forming feedforward and feedback of quantum input-output systems has
been developed recently where the concept of the series product for
determining the generators of systems in series was introduced \cite{GJ
Series}. This is part of a more general theory which extends to (indirect)
feedback loops mediated through beam splitters \cite{QFN1}.

In this paper, we wish to describe a coherent control strategy which allows
for a \textit{feedback cancellation} of the stochastic component of the
dynamics. Here the input noise is passed through the system, has a signal
added, is passed through the system again to undo the stochastic component
of the evolution, and finally has the original signal subtracted. Assuming
that these operations occur without time delays, the overall result is a 
\textit{closed} system dynamics with a modified Hamiltonian which \textit{%
generically} is bilinear.

\section{Input-Output Devices}

In a Markov model of an open quantum mechanical system interacting driven by
a bosonic field in the vacuum state, we introduce input processes $%
b_{i}\left( t\right) $ for $i=1,\cdots ,n$ with the canonical commutation
relations, \cite{Gardiner Collett}, 
\begin{equation}
\lbrack b_{i}\left( t\right) ,b_{j}^{\dag }\left( s\right) ]=\delta
_{ij}\,\delta \left( t-s\right) .
\end{equation}
It is convenient to write these as a column vector of length $n$ 
\begin{equation}
\mathbf{b}^{\text{in}}\left( t\right) =\left( 
\begin{array}{c}
b_{1}\left( 1\right) \\ 
\vdots \\ 
b_{n}\left( t\right)
\end{array}
\right) .
\end{equation}
We sketch the system plus field as a two port device having an input and an
output port.

\begin{center}
\setlength{\unitlength}{.04cm}
\begin{picture}(120,45)
\label{pic1}
\thicklines
\put(45,10){\line(0,1){20}}
\put(45,10){\line(1,0){30}}
\put(75,10){\line(0,1){20}}
\put(45,30){\line(1,0){30}}
\thinlines
\put(48,20){\vector(-1,0){45}}
\put(120,20){\vector(-1,0){20}}
\put(120,20){\line(-1,0){48}}
\put(50,20){\circle{4}}
\put(70,20){\circle{4}}
\put(100,26){input, ${\bf b}^{\rm in}$}
\put(32,39){system $(S,{\bf L},H)$}
\put(-20,26){output, ${\bf b}^{\rm out}$}
\end{picture}
%

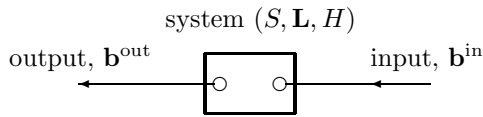
Figure 1: input-output component
\end{center}

The evolution can be described by the unitary process $\left\{ V_{t}\right\} 
$ satisfying the Wick-ordered (creators appear on the left, annihilators on
the right) differential equation of the form 
\begin{eqnarray*}
\frac{d}{dt}V_{t} &=&\mathbf{b}^{\text{in}}\left( t\right) ^{\dag }(S\left(
t\right) -I)V_{t}\mathbf{b}^{\text{in}}\left( t\right) +\mathbf{b}^{\text{in}%
}\left( t\right) ^{\dag }\mathbf{L}\left( t\right) V_{t} \\
&&-\mathbf{L}^{\dag }\left( t\right) S\left( t\right) V_{t}\mathbf{b}^{\text{%
in}}\left( t\right) -(\frac{1}{2}\mathbf{L}^{\dag }\left( t\right) \mathbf{L}%
\left( t\right) -iH\left( t\right) )V_{t},
\end{eqnarray*}
with initial value $V_{0}=I$. The coefficients $S_{ij}\left( t\right) $, $%
L_{i}\left( t\right) $ and $H\left( t\right) $ are operators having the
property that they are adapted (i.e., commute with the fields $b_{j}\left(
s\right) $ and $b_{j}^{\dag }\left( s\right) $ for earlier times $s<t$, and
such that $S\left( t\right) =\left( S_{ij}\left( t\right) \right) $ is a
unitary matrix with operator entries (i.e., $\sum_{k}S_{ik}S_{jk}^{\dag
}=\delta _{ij}=\sum_{k}S_{ki}^{\dag }S_{kj}$), $\mathbf{L}\left( t\right) $
in the column vector of operators $\left( L_{i}\left( t\right) \right) $,
and $H$ is self-adjoint. This equation can be interpreted as a quantum
stochastic differential equation \cite{Gardiner Collett}, \cite{HP}. The
solution is known to exist and be unique, and we shall denote it as 
\begin{equation}
V_{t}=V_{t}\left( S,\mathbf{L},H;\mathbf{b}^{\text{in}}\right)
\end{equation}

Let $X$ be a fixed operator of the system and set $\tilde{X}\left( t\right)
\triangleq V_{t}{}^{\dag }XV_{t}$, then we obtain the Heisenberg-Langevin
equation 
\begin{multline}
\frac{d}{dt}\tilde{X}\left( t\right) =\mathcal{V}_{t}\left( X;S,\mathbf{L},H;%
\mathbf{b}^{\text{in}}\right)  \notag \\
\equiv \mathbf{b}^{\text{in}}\left( t\right) ^{\dag }\mathcal{S}\left(
X;t\right) \mathbf{b}^{\text{in}}\left( t\right) +\mathbf{b}^{\text{in}%
}\left( t\right) ^{\dag }\mathcal{J}\left( X,t\right) +\mathcal{K}\left(
X,t\right) \mathbf{b}^{\text{in}}\left( t\right) +\mathcal{L}\left(
X,t\right) ,  \label{Heis-Lang}
\end{multline}
where we encounter the super-operators 
\begin{eqnarray*}
\mathcal{S}\left( X;t\right) _{ij} &=&\sum_{k}\tilde{S}_{ki}^{\dag }\tilde{X}%
\tilde{S}_{kj}-\tilde{X}\delta _{ij}, \\
\mathcal{J}\left( X,t\right) _{i} &=&\sum_{j}\tilde{S}_{ji}^{\dag }[\tilde{X}%
,\tilde{L}_{j}],\ \mathcal{K}\left( X,t\right) _{j}=\sum_{i}[\tilde{L}%
_{i}^{\dag },X]\tilde{S}_{ij}, \\
\mathcal{L}\left( X,t\right) &=&\frac{1}{2}\sum_{i}\tilde{L}_{i}^{\dag }[%
\tilde{X},\tilde{L}_{i}]+\frac{1}{2}\sum_{i}[\tilde{L}_{i}^{\dag },\tilde{X}]%
\tilde{L}_{i}-i[\tilde{X},\tilde{H}].
\end{eqnarray*}
Here $\tilde{S}_{ij}\left( t\right) $ denotes $V_{t}^{\dag }S_{ij}\left(
t\right) V_{t}$, etc. The final term is a (time-dependent) Lindbladian. In
the case $S=1$, this equation reduces to the Heisenberg-Langevin equations
introduced by Gardiner and Collett \cite{Gardiner Collett}. The output
fields are defined by $b_{i}^{\text{out}}\left( t\right) =V_{t}^{\dag
}b_{i}\left( t\right) V_{t}$ and we have the input-output relation 
\begin{equation}
\mathbf{b}^{\text{out}}\left( t\right) =\tilde{S}\left( t\right) \mathbf{b}^{%
\text{in}}\left( t\right) +\mathbf{\tilde{L}}\left( t\right) ,
\end{equation}
that is, $b_{i}^{\text{out}}\left( t\right) =\sum_{j=1}^{n}V_{t}^{\dag
}S_{ij}\left( t\right) V_{t}b_{j}\left( t\right) +V_{t}^{\dag }L_{i}\left(
t\right) V_{t}$.

\subsection{Examples}

\subsubsection{Cavity Mode}

For a cavity mode $a$ and complex damping $\kappa =\frac{1}{2}\gamma
+i\omega $ with a single input field, we have 
\begin{equation*}
V_{t}=V_{t}\left( I,\sqrt{\gamma }a,\omega a^{\dag }a;b^{\text{in}}\right) .
\end{equation*}
Here $b^{\text{out}}\left( t\right) =b^{\text{in}}\left( t\right) +\sqrt{%
\gamma }V_{t}^{\dag }aV_{t}$.

\subsubsection{Pure Hamiltonian, $\mathtt{HAM}\left( H\right) $}

The situation of a closed system is described by the device $\mathtt{HAM}%
\left( H\right) $, 
\begin{equation*}
V_{t}=V_{t}\left( I,0,H;\mathbf{b}^{\text{in}}\right)
\end{equation*}
and there is no interaction between the system and the input fields. For $H$
time-independent, we have $V_{t}\equiv \exp \left\{ -itH\right\} $.

\subsubsection{Beam Splitter, $\mathtt{BS}\left( T\right) $}

We take\ $S=T$ a unitary matrix with c-number entries. The beam splitter
with matrix $T$ is the device $\mathtt{BS}\left( T\right) $ described by 
\begin{equation*}
V_{t}=V_{t}\left( T,0,0;\mathbf{b}^{\text{in}}\right) .
\end{equation*}
The system dynamics is trivial nd we have the input-output relation 
\begin{equation}
\mathbf{b}^{\text{out}}\left( t\right) =T\mathbf{b}^{\text{in}}\left(
t\right) .
\end{equation}

\subsubsection{Signal Adding Devices, $\mathtt{ADD}\left( \mathbf{u}\right) $%
}

Let $\mathbf{u}=\left( u_{j}\right) $ be a square-integrable function of
time taking values in $\mathbb{C}^{n}$. We consider the device with dynamics 
\begin{equation*}
V_{t}=V_{t}\left( I,\mathbf{u},0;\mathbf{b}^{\text{in}}\right) .
\end{equation*}
This has trivial system dynamics and input-output relation 
\begin{equation*}
\mathbf{b}^{\text{out}}\left( t\right) =\mathbf{b}^{\text{in}}\left(
t\right) +\mathbf{u}\left( t\right) .
\end{equation*}
Here we think of $\mathbf{u}$ as a signal carried by the input field.
Alternatively we may think of the field as now being in the coherent state
with intensity $\mathbf{u}$. We shall refer to such a component \textit{%
device adding a signal} $\mathbf{u}$ to the field. Such a device will be
denoted as $\mathtt{ADD}\left( \mathbf{u}\right) $.

\section{Systems in Series}

Let us consider two systems in cascade as shown below.

\begin{center}
%
%
%
%
%
%
%
%
%
\setlength{\unitlength}{.1cm}
\begin{picture}(72,22)
\label{pic2}

\thicklines

\put(10,5){\line(0,1){10}}
\put(10,5){\line(1,0){20}}
\put(30,5){\line(0,1){10}}
\put(10,15){\line(1,0){20}}

\put(40,5){\line(0,1){10}}
\put(40,5){\line(1,0){20}}
\put(60,5){\line(0,1){10}}
\put(40,15){\line(1,0){20}}

\thinlines
\put(12,10){\vector(-1,0){15}}
\put(42,10){\vector(-1,0){14}}

\put(72,10){\vector(-1,0){14}}

\put(13,10){\circle{2}}
\put(43,10){\circle{2}}

\put(27,10){\circle{2}}
\put(57,10){\circle{2}}

\put(11,18){$(S_2 , {\bf L}_2 , H_2)$}
\put(40,18){$(S_1 , {\bf L}_1 , H_1)$}

\end{picture}%
%

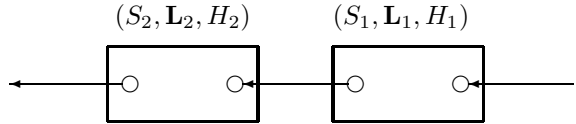
figure 2: Cascaded systems
\end{center}

The time delay for the output of the first system $\left( S_{1},\mathbf{L}%
_{1},H_{1}\right) $ to reach the second system $\left( S_{2},\mathbf{L}%
_{2},H_{2}\right) $ as input will be some $\tau >0$, and we are interested
in the limit situation where $\tau \rightarrow 0$. This leads to a single
effective Markovian model with coefficients given by the \textit{series
product} \cite{GJ Series} 
\begin{equation}
\left( S_{2},\mathbf{L}_{2},H_{2}\right) \vartriangleleft \left( S_{1},%
\mathbf{L}_{1},H_{1}\right) \triangleq \left( S_{2}S_{1},\mathbf{L}_{2}+S_{2}%
\mathbf{L}_{1},H_{1}+H_{2}+\mathrm{Im}\mathbf{L}_{2}^{\dag }S_{2}\mathbf{L}%
_{1}\right) .
\end{equation}
We remark that the series product is valid even when the observables of the
two systems are not suppose to commute, that is, we have feedback into the
same system. Figure 3 below shows a direct feedback situation with an
equivalent model given by the series product.

\begin{center}
%
%
%
%
%
%
%
%
%
\setlength{\unitlength}{.04cm}
\begin{picture}(100,40)
\label{pic3a}

\put(63,37){$1$}
\put(63,18){$2$}

\thicklines
\put(45,15){\line(0,1){30}}
\put(45,15){\line(1,0){30}}
\put(75,15){\line(0,1){30}}
\put(45,45){\line(1,0){30}}

\thinlines
\put(15,35){\vector(1,0){15}} 
\put(15,35){\line(1,0){70}} 
\put(60,35){\circle*{4}} 
\put(85,35){\line(0,-1){8}}
\put(85,23){\line(0,-1){23}}
\put(15,0){\line(1,0){70}}
\put(15,0){\line(0,1){25}}

\put(15,25){\vector(1,0){15}} 
\put(15,25){\vector(1,0){90}} 
\put(60,25){\circle*{4}} 

\put(94,29){${\bf b}^{\rm out}$}
\put(12,38){${\bf b}^{\rm in}$}

\end{picture}
%
%
%
%
%
%
%
%
%
%
\setlength{\unitlength}{.04cm}
\begin{picture}(80,40)
\label{pic3b}

\thinlines
\put(63,37){$1$}
\put(53,18){$2$}

\thicklines
\put(45,15){\line(0,1){30}}
\put(45,15){\line(1,0){30}}
\put(75,15){\line(0,1){30}}
\put(45,45){\line(1,0){30}}

\thinlines
\put(15,35){\vector(1,0){15}} 
\put(15,35){\line(1,0){45}} 
\put(60,35){\circle*{4}} 
\put(60,25){\line(0,1){10}}

\put(60,25){\vector(1,0){35}} 
\put(60,25){\circle*{4}}

\put(82,30){${\bf b}^{\rm out}$}
\put(14,38){${\bf b}^{\rm in}$}

\end{picture}
%

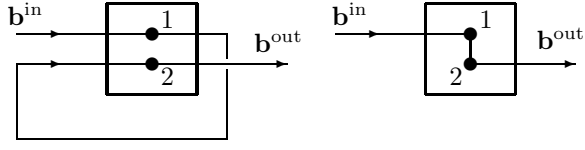
Figure 3: Direct feedback, $S_{2}\triangleleft S_{1}$.
\end{center}

The series product is generally not commutative, but is associative. Pure
Hamiltonian devices can be entered in series at any point due to the
identity 
\begin{equation}
\left( S,\mathbf{L},H\right) \vartriangleleft \left( I,0,H_{0}\right)
=\left( S,\mathbf{L},H+H_{0}\right) =\left( I,0,H_{0}\right)
\vartriangleleft \left( S,\mathbf{L},H\right) .  \label{series ham com}
\end{equation}

\subsection{Derivation of the Series Product}

A rigorous derivation of the series product is given in \cite{GJ Series}, 
\cite{QFN1}, however, we now give an alternative heuristic derivation
extending the argument originally used by Gardiner \cite{Gardiner93}. For
convenience we work withonly one noise field and supress the time variable.
The Heisenberg-Langevin equation for any observable $X$ of the joint system
will be 
\begin{equation*}
\frac{d}{dt}\tilde{X}\left( t\right) =\sum_{\alpha =1,2}\mathcal{V}%
_{t}\left( X;S_{\alpha },L_{\alpha },H_{\alpha };b_{\alpha }^{\text{in}%
}\right)
\end{equation*}
where $b_{1}^{\text{in}}\left( t\right) $ is the overall input $b^{\text{in}%
}\left( t\right) $ and 
\begin{equation*}
b_{2}^{\text{in}}\left( t\right) =b_{1}^{\text{out}}\left( t-0^{+}\right)
\equiv \tilde{S}_{1}\left( t\right) b^{\text{in}}\left( t\right) +\tilde{L}%
_{1}\left( t\right) .
\end{equation*}
The super-operators correspond to the coefficients for the first and second
system respectively for $\alpha =1,2$. We eliminate the fields $b_{\alpha }^{%
\text{in}}\left( t\right) $ and write in terms of $b^{\text{in}}\left(
t\right) $. After some algebra we find that the Langevin equation has the
form $\left( \ref{Heis-Lang}\right) $ with 
\begin{eqnarray*}
\mathcal{S}\left( X\right) &=&\mathcal{S}_{1}\left( X\right) +\tilde{S}%
_{1}^{\dag }\mathcal{S}_{2}\left( X\right) \tilde{S}_{1}\equiv (\tilde{S}_{2}%
\tilde{S}_{1}^{\dag })\tilde{X}(\tilde{S}_{2}\tilde{S}_{1})-\tilde{X} \\
\mathcal{J}\left( X\right) &=&\mathcal{J}_{1}\left( X\right) +\tilde{S}%
_{1}^{\dag }\mathcal{J}_{2}\left( X\right) +\tilde{S}_{1}^{\dag }\mathcal{S}%
_{2}\left( X\right) \tilde{L}_{1}\equiv (\tilde{S}_{2}\tilde{S}_{1})^{\dag }%
\left[ \tilde{X},\tilde{L}_{2}+\tilde{S}_{2}\tilde{L}_{1}\right] , \\
\mathcal{K}\left( X\right) &=&\mathcal{K}_{1}\left( X\right) +\mathcal{K}%
_{2}\left( X\right) \tilde{S}_{2}+\tilde{L}_{1}^{\dag }\mathcal{S}_{2}\left(
X\right) \tilde{S}\equiv \lbrack (\tilde{L}_{2}+\tilde{S}_{2}\tilde{L}%
_{1})^{\dag },\tilde{X}](\tilde{S}_{2}\tilde{S}_{1}) \\
\mathcal{L}\left( X\right) &=&\mathcal{L}_{1}\left( X\right) +\mathcal{L}%
_{2}\left( X\right) +\tilde{L}_{1}^{\dag }\mathcal{J}_{2}\left( X\right) +%
\mathcal{K}_{2}\left( X\right) \tilde{L}_{1}+\tilde{L}_{1}^{\dag }\mathcal{S}%
_{2}\left( X\right) \tilde{L}_{1}
\end{eqnarray*}
and by inspection we deduce that this is of the standard form with $S\equiv
S_{2}S_{1}$, $L\equiv L_{2}+S_{2}L_{1}$. After some algebra we show that $%
\mathcal{L}\left( X\right) $ is a Lindbladian with coupling operator $\tilde{%
L}$ and Hamiltonian $\tilde{H}\equiv $\textrm{Im}$\tilde{L}_{2}^{\dag }%
\tilde{S}_{2}\tilde{L}_{1}$.

\subsection{Pre- and Post-applications of beam splitters}

Given a fixed component with coefficients $\left( S,\mathbf{L},H\right) $,
we apply a beam splitter $T$ to the inputs prior to entry and the inverse $%
T^{-1}$ to outputs after exit from the component. Assuming zero time delay
in travelling from the beam splitters to the components, we have for the
combined system 
\begin{equation}
\left( T^{-1},0,0\right) \vartriangleleft \left( S,\mathbf{L},H\right)
\vartriangleleft \left( T,0,0\right) =\left( T^{-1}ST,T^{-1}\mathbf{L}%
,H\right) .  \label{trick}
\end{equation}
We note the identity 
\begin{equation*}
V_{t}\left( S,\mathbf{L},H;T\mathbf{b}^{\text{in}}\right) \equiv V_{t}\left(
T^{-1}ST,T^{-1}\mathbf{L},H;\mathbf{b}^{\text{in}}\right) ,
\end{equation*}
which means that the combined system could be viewed as the original system,
but driven by ``rotated'' input $T\mathbf{b}^{\text{in}}$.

\section{Bilinear Hamiltonians}

\subsection{Constructions with Noise}

Let us fix a matrix $T$ of c-numbers. We pass the initial input through a
device adding a signal $\mathbf{v}=T^{-1}\mathbf{u}$, and then pass the
output as input through a general component with coefficients $\left( T,%
\mathbf{L},H\right) $. In the zero-delay limit, the combined system is
determined by 
\begin{equation*}
\left( T,\mathbf{L},H_{0}\right) \vartriangleleft \left( I,T^{-1}\mathbf{u}%
,0\right) =\left( T,\mathbf{L}+\mathbf{u},H_{0}+\mathrm{Im}\mathbf{L}^{\dag }%
\mathbf{u}\right) .
\end{equation*}
We may subsequently pass the output through a third component which adds the
signal $-\mathbf{u}$, we find 
\begin{equation*}
\left( I,-\mathbf{u},0\right) \vartriangleleft \left( T,\mathbf{L}%
,H_{0}\right) \vartriangleleft \left( I,T^{-1}\mathbf{u},0\right) =\left( T,%
\mathbf{L},H_{0}+2\mathrm{Im}\mathbf{L}^{\dag }\mathbf{u}\right) .
\end{equation*}
The result is that we modify the Hamiltonian to 
\begin{equation}
H\left( \mathbf{u}\right) =H_{0}+2\mathrm{Im}\mathbf{L}^{\dag }\mathbf{u}%
=H_{0}+2\sum_{j}\left( L_{j,R}u_{j,I}-L_{j,I}u_{j,R}\right)
\end{equation}
where $u_{j}=u_{j,R}+iu_{j,I}$ and $L_{j}=L_{j,R}+iL_{j,I}$ with $%
u_{j,R},u_{j,I}$ real and $L_{j,R},L_{j,I}$ self-adjoint.

The combined model includes a coupling via the operator $\mathbf{L}$ to the
environment so that the evolution, given by $V_{t}\left( T,\mathbf{L}%
,H\left( \mathbf{u}\right) ,\mathbf{b}^{\text{in}}\right) $. To counteract
this now stochastic evolution we may perform a homodyne measurement on a
quadrature of $\mathbf{b}^{\text{out}}$ and employ a filter to estimate the
quantum state of the system conditioned on the measurement record \cite
{Belavkin83}.

\subsection{Constructions without Noise}

We now show how, given a fixed system with internal Hamiltonian $H_{0}$ and
which can couple to input noise with operators $\mathbf{L}$, to construct an
effectively closed system with internal Hamiltonian $H\left( \mathbf{u}%
\left( t\right) \right) $ for prescribed signal $\mathbf{u}$, with bilinear
Hamiltonian $H\left( \mathbf{u}\right) =H_{0}+$\textrm{Im}$\mathbf{L}^{\dag }%
\mathbf{u}$. We refer to this general procedure as \textit{quantum noise
cancellation by feedback}. The system has been engineered so that it
interacts with the input noise, but this is then undone by feedback of the
output noise back into the system. Between these events, we shall add in the
signal term which we later subtract out.

Let \texttt{SYS}$\left( \mathbf{L}\right) $ denote the device with
coefficients $\left( I,\mathbf{L},0\right) $, then what we shall show is
that we can obtain the closed system $\mathtt{HAM}\left( H\left( \mathbf{u}%
\right) \right) $ from only the devices $\mathtt{SYS}\left( \mathbf{L}%
\right) ,\mathtt{ADD}\left( \pm \mathbf{u}\right) $, $\mathtt{BS}\left(
-I\right) ,$ and $\mathtt{HAM}\left( H_{0}\right) $ in series.

Using identity $\left( \ref{trick}\right) $ with $T=-I$ we have 
\begin{equation}
\left( -I,0,0\right) \vartriangleleft \left( I,\mathbf{L},0\right)
\vartriangleleft \left( -I,0,0\right) =\left( I,-\mathbf{L},0\right) ,
\label{reversal}
\end{equation}
or \texttt{BS}$\left( -I\right) \vartriangleleft $\texttt{SYS}$\left( 
\mathbf{L}\right) \vartriangleleft $\texttt{BS}$\left( -I\right) =$\texttt{%
SYS}$\left( -\mathbf{L}\right) $.

Now \texttt{ADD}$\left( \mathbf{u}\right) \vartriangleleft $\texttt{SYS}$%
\left( -\mathbf{L}\right) \vartriangleleft $\texttt{ADD}$\left( -\mathbf{u}%
\right) =$\texttt{SYS}$\left( \mathbf{L}\right) $, that is, 
\begin{equation*}
\left( I,\mathbf{u},0\right) \vartriangleleft \left( I,-\mathbf{L},0\right)
\vartriangleleft \left( I,-\mathbf{u},0\right) =\left( I,-\mathbf{L},\mathrm{%
Im}\mathbf{L}^{\dag }\mathbf{u}\right)
\end{equation*}
and this can be used to exactly negate the $\mathbf{L}$-coupling: 
\begin{equation*}
\left( I,-\mathbf{L},\mathrm{Im}\mathbf{L}^{\dag }\mathbf{u}\right)
\vartriangleleft \left( I,\mathbf{L},0\right) =\left( I,0,\mathrm{Im}\mathbf{%
L}^{\dag }\mathbf{u}\right) .
\end{equation*}
The sequence of output-to-input connections is then given by 
\begin{equation*}
\mathtt{ADD}\left( \mathbf{u}\right) \vartriangleleft \mathtt{BS}\left(
-I\right) \vartriangleleft \mathtt{SYS}\left( \mathbf{L}\right)
\vartriangleleft \mathtt{BS}\left( -I\right) \vartriangleleft \mathtt{ADD}%
\left( -\mathbf{u}\right) \vartriangleleft \mathtt{SYS}\left( \mathbf{L}%
\right) =\mathtt{HAM}\left( \mathrm{Im}\mathbf{L}^{\dag }\mathbf{u}\right) .
\end{equation*}
From $\left( \ref{series ham com}\right) $ we may include the pure
Hamiltonian device $\mathtt{HAM}\left( H_{0}\right) $ at any stage in the
series, in particular, $\mathtt{HAM}\left( H_{0}\right) \vartriangleleft 
\mathtt{HAM}\left( \mathrm{Im}\mathbf{L}^{\dag }\mathbf{u}\right) =\mathtt{%
HAM}\left( H_{0}+\mathrm{Im}\mathbf{L}^{\dag }\mathbf{u}\right) $.

The construction requires the introduction of noise through coupling to the
input field, but relies on a cancellation of the stochastic component of the
dynamics by passing the output field through the system a second time. To
achieve the cancellation, we needed to reverse the sign of the coupling
operators $\mathbf{L}$ which is achieved by $\left( \ref{reversal}\right) $.
As the noise is being passed through the same physical system twice, we have
an example of feedback, however, as we have mentioned, the series product
covers this situation.

\subsection{Physical Example}

As a possible model application we can consider the all-optical feedback
experiment proposed by Wiseman and Milburn, \cite{WM} section II.B. Here a
cavity mode with annihilator $a$ is contained between two mirrors. The input
is an external light beam which is shone on the first mirror where it is
reflected. The reflection of the mirror induces an interaction with the
cavity mode $\left( L=\sqrt{\gamma }a\right) $ and is the reflected beam
picks up a phase $e^{i\theta }$. The reflected beam is thenshone on the
second mirror again interacting with the cavity mode in the same way. The
series product has been previously used to rederive model specification of
the resulting model \cite{GJ Series}, section IV.A.

We now propose a modification where the signal is added to the light beam on
route from being reflected off the first mirror to impinging on the second,
and then subtracted from the light reflected from the second mirror. We
additionally assume that $\theta =\pi $ so that when the beam is relected
off either cavity mirror it picks up a sign change. Suppose that the
internal Hamiltonian is $H_{0}=\omega _{0}a^{\dag }a$ and $a\left( t\right)
=V_{t}^{\dag }aV_{t}$ is the mode in the Heisenberg picture, then the
Heisenberg-Langevin equation will be 
\begin{equation*}
\dot{a}\left( t\right) =-i\omega _{0}a\left( t\right) -\frac{\sqrt{\gamma }}{%
2}u\left( t\right)
\end{equation*}
corresponding to evolution under the Hamiltonian $\omega _{0}a^{\dag }a+%
\frac{\sqrt{\gamma }}{2i}\left( a^{\dag }u\left( t\right) -au^{\ast }\left(
t\right) \right) $. This is the familiar situation of a driven harmonic
oscillator with control function $u$ \cite{HeffnerLouisell} where it is
known that the vacuum may only be steered to a coherent state.


\begin{thebibliography}{99}
\bibitem{D'Alessandro}  D. D'Alessandro, Introduction to Quantum Control and
Dynamics, Chapman \& Hall/CRC, 2008

\bibitem{BRB}  R.W. Brockett, C. Rangan, A.M. Bloch, The controllability of
infinite quantum systems, in Proceedings of the 42-th IEEE Conference on
Decision and Control, 4-7 Dec., 428-433, 2003

\bibitem{SchP-RW}  S.G. Schirmer, P.J. Pemberton-Ross, X. Wang, Comparative
Analysis of Control Strategies, Proceedings of PhysCon2007, IPACS Electronic
Library: arXiv:0801.0746v1 [quant-ph]

\bibitem{Sch01}  S.G. Schirmer, Quantum control using Lie group
decompositions, in Proceedings of the 40-th IEEE Conference on Decision and
Control, 4-7 Dec., 293-303, vol. 1, 2001

\bibitem{Belavkin83}  V. P. Belavkin: Theory of the Control of Observable
Quantum Systems. Automatica and Remote Control \textbf{44} (2) 178--188
(1983).

\bibitem{BvH}  L. Bouten, R. van Handel, On the separation principle of
quantum control, arXiv:math-ph/0511021

\bibitem{Lloyd}  S. Lloyd, \textit{Coherent quantum control}, Phys. Rev. A,
62:022108, 2000

\bibitem{JNP}  M. R. James, H. I. Nurdin, and I. R. Petersen, \textit{H}$%
^{\infty }$\textit{\ control of linear quantum stochastic systems}, 2007, to
appear in IEEE Transactions on Automatic Control. http://arxiv:
quant-ph/0703150

\bibitem{Mabuchi}  H. Mabuchi, \textit{Coherent-feedback control with a
dynamic compensator}, March 2008, submitted for publication, preprint:
http://arxiv.org/abs/0803.2007.

\bibitem{Gardiner Collett}  C.W. Gardiner and M.J. Collett. Input and output
in damped quantum systems: Quantum stochastic differential equations and the
master equation. Phys. Rev. A, \textbf{31}(6):37613774, (1985)

\bibitem{GJ Series}  J. Gough, M.R. James, \textit{The series product and
its application to feedforward and feedback networks}, arXiv:07070048(v1)
[quant-ph]

\bibitem{QFN1}  J. Gough, M.R. James, \textit{Quantum\ Feedback Networks:
Hamiltonian Formulation}, arXiv:07070048(v1) [quant-ph]

\bibitem{HP}  R. L. Hudson and K. R. Parthasarathy, \textit{Quantum Ito's
formula and stochastic evolutions,} Commun. Math. Phys. \textbf{93}, 301-323
(1984)

\bibitem{WM}  H. M. Wiseman and G. J. Milburn. All-optical versus
electro-optical quantum-limited feedback. Phys. Rev. A, 49(5):41104125, 1994.

\bibitem{Gardiner93}  C.W. Gardiner. Driving a quantum system with the
output field from another driven quantum system. Phys. Rev. Lett., \textbf{70%
}(15):22692272, 1993.

\bibitem{HeffnerLouisell}  H. Heffner, W.H. Louisell, J. Math Phys., \textbf{%
6}, 474 (1965)
\end{thebibliography}
\end{document}